\documentclass[modern]{aastex631}

\usepackage{hyperref}
\usepackage{breakurl} 

\shorttitle{TrExoLiSTS}
\shortauthors{Nikolov et al.}

\begin{document}

\title{TrExoLiSTS: Transiting Exoplanets List of Space Telescope Spectroscopy}

\correspondingauthor{Nikolay K. Nikolov}
\email{nnikolov@stsci.edu}

\author[0000-0002-6500-3574]{Nikolay K. Nikolov}
\affiliation{Space Telescope Science Institute, 3700 San Martin Drive, Baltimore, MD 21218, USA}

\author{Aiden Kovacs}
\affiliation{Space Telescope Science Institute, 3700 San Martin Drive, Baltimore, MD 21218, USA}

\author{Catherine Martlin}
\affiliation{Space Telescope Science Institute, 3700 San Martin Drive, Baltimore, MD 21218, USA}




\begin{abstract}
We present the STScI WFC3 project webpage, Transiting Exoplanets List of Space Telescope Spectroscopy, TrExoLiSTS. It tabulates existing observations of transiting exoplanet atmospheres, available in the MAST archive made with {\it{HST}} WFC3 using the stare or spatial scan mode. A parallel page is available for all instruments aboard {\it{JWST}} using the spectral Time Series Observation (TSO) mode. The webpages include observations obtained during primary transits, secondary eclipses and phase curves. TREXOLISTS facilitates proposal preparation for programs that are highly-complementary to existing programs in terms of targets, wavelength coverage, as well as reduces duplication and redundant effort. Reference for the quality of the {\it{HST}} WFC3 visits taken more than 1.5 years ago are made available via including diagrams of the direct image, white light curve and drift of the spectral time series across the detector. Future improvements to the webpage will include: Expanding program query to other {\it{HST}} instruments and reference for the quality of {\it{JWST}} visits.

\end{abstract}

\keywords{Catalogs(205) --- Exoplanet catalogs(488) --- Exoplanet atmospheres(487) --- Observational astronomy(1145)}


\section{Introduction} \label{sec:intro}
Over the past two decades, spectroscopic observations of transiting exoplanet atmospheres with the {\it{Hubble Space Telescope}} ({\it{HST}}) have been steadily increasing, covering primary transits, secondary eclipses and phase curves. Thanks to the availability of a drift scanning mode, the infrared channel of the Wide Field Camera 3 (WFC3) has become a leading instrument capable of delivering high signal to noise observations sensitive to absorption from water in exoplanet atmospheres. With multiple community efforts aimed toward establishing a diverse and statistically large sample of compositional constraints of exoplanet atmospheres, a need for a catalog of archived {\it{HST}} observations has emerged. To address this need and help reduce target duplication and redundant effort in proposal preparation, the WFC3 team has established a webpage, TrExoLiSTS: Transiting Exoplanets List of Space Telescope Spectroscopy. It provides a summary of times series observations of transiting exoplanets that are archived on the MAST\footnote{\url{https://archive.stsci.edu/}}. TREXOLISTS is organized in the form of a table and is equipped with a search bar allowing quick identification of targets by name or coordinates, program identifier e.g., GO, DD; PI name, and can be accessed at \url{https://www.stsci.edu/~WFC3/trexolists/trexolists.html}. With the launch of the {\it{James Webb Space Telescope}} ({\it{JWST}}) and first observations of transiting exoplanets, TREXOLISTS has been expanded and includes archived observations made using the Time Series Observing (TSO) mode with each instrument of the {\it{JWST}} and is available as a separate webpage at \url{https://www.stsci.edu/~nnikolov/TrExoLiSTS/JWST/trexolists.html}.

\section{Methods} \label{sec:style}
The contents of TREXOLISTS is produced using public {\it{HST}} and {\it{JWST}} program information provided by the Space Telescope Science Institute (STScI). Lists of programs are obtained from the following sources: 

\begin{itemize}
 \item{{\it{HST}} TAC Programs:  \url{https://www.stsci.edu/ftp/presto/ops/program-lists/HST-TAC.html}}
 \item{{\it{HST}} Mid-Cycle Approved Programs: \url{https://hst-docs.stsci.edu/hsp/hst-mid-cycle-approved-programs}}
 \item{{\it{HST}} Director's Discretionary Programs: \url{https://www.stsci.edu/ftp/presto/ops/program-lists/HST-DD.html}}
 \item{{\it{JWST}} First Image Observations: \url{https://www.stsci.edu/jwst/science-execution/approved-programs/webb-first-image-observations}}
 \item{{\it{JWST}} Guaranteed Time Observation Programs: \url{https://www.stsci.edu/jwst/science-execution/approved-programs/guaranteed-time-observations}}
 \item{{\it{JWST}} Director’s Discretionary Early Release Science Programs: \url{https://www.stsci.edu/jwst/science-execution/approved-ers-programs}}
 \item{{\it{JWST}} General Observer Programs in Cycle 1: \url{https://www.stsci.edu/jwst/science-execution/approved-programs/cycle-1-go}}
\end{itemize}

In addition to the listed sources, we also included {\it{JWST}} commissioning programs 1541 (NIRISS, PI: Espinoza), 1118 (NIRSpec, PI: Proffitt), 1442 (NIRCam, PI: Schlawin) and 1033 (MIRI, PI: Kendrew). 

A custom code has been developed to identify programs that include time series observations of transiting exoplanets to parse target and observing parameters. To achieve these goals, the code first downloads the public submissions made via the Astronomers Proposal Tool (APT, \citealt{2004SPIE.5493..351R}) along with the visit files that describe the status of the observations. Identification of appropriate programs has been performed by a search for key words in their APT submission. A future improvement of this step is envisioned to include a Machine Learning algorithm that identifies potential programs. All selected programs are then parsed following the characteristics in the structure of the APT files for each observatory. Information regarding target coordinates, instrument settings, and observing details are obtained and saved in a table, which is provided on the TREXOLISTS webpage in machine readable format. 

TREXOLISTS content is structured as follows:

\begin{itemize}
 \item{Program identifier hyperlinked with the relevant full information provided by STScI for the observing program}
 \item{Target name, including alternative names\footnote{When available in the APT submission \label{fn1}} hyperlinked to exo.MAST \citep{Mullally_2019}}
 \item{Right Ascension and Declination} 
 \item{Instrument: UVIS or IR for {\it{HST}} WFC3 and NIRCam, NIRISS, NIRSpec and MIRI for {\it{JWST}}, respectively}
 \item{Dispersive Element for {\it{HST}} WFC3: G280, G102 or G141; or Mode for {\it{JWST}}: SOSS, BOTS+PRISM, GRISMR+F322W2, GRISMR+F444W, LRS, MRS, etc.}
\item{Phase Start/End, planet phase between which observation has started}
\item{Observing mode, stare or scan}
\item{Scan rate in arc-sec per seconds}
\item{Visit identifier}
\item{Date/Time Start/End, date and time of the observation start/end}
\item{Orbits for {\it{HST}} and Duration in hours for {\it{JWST}}, obtained from the APT and visit files, respectively}
\item{Status, Archived or Failed}; scheduling window for upcoming observations is envisioned to be included
\item{PI family name}
\item{Proprietary Period in Months}
\end{itemize}

In addition to the listed parameters, three additional columns provide hyperlinks to WFC3 Direct Image or {\it{JWST}} Target Acquisition, White Light Curve and XY drift map of the spectra on the detector. These products are currently available only for {\it{HST}} WFC3 IR observations via products from the WFC3's Quicklook project \citep{2017IAUS..325..397B, 2019wfc..rept...12S}. Placeholders have been included for {\it{JWST}} envisioning a future inclusion of such equivalent products for {\it{JWST}} observations. TREXOLISTS includes a user-friendly interface for easier navigation through the content and a search bar allowing quick identification of targets by name or equatorial coordinates, program identifiers, instrument settings, observing dates, number of {\it{HST}} orbits, durations of {\it{JWST}} observations and visit identifiers.

\begin{figure}
\epsscale{1.1}
\plotone{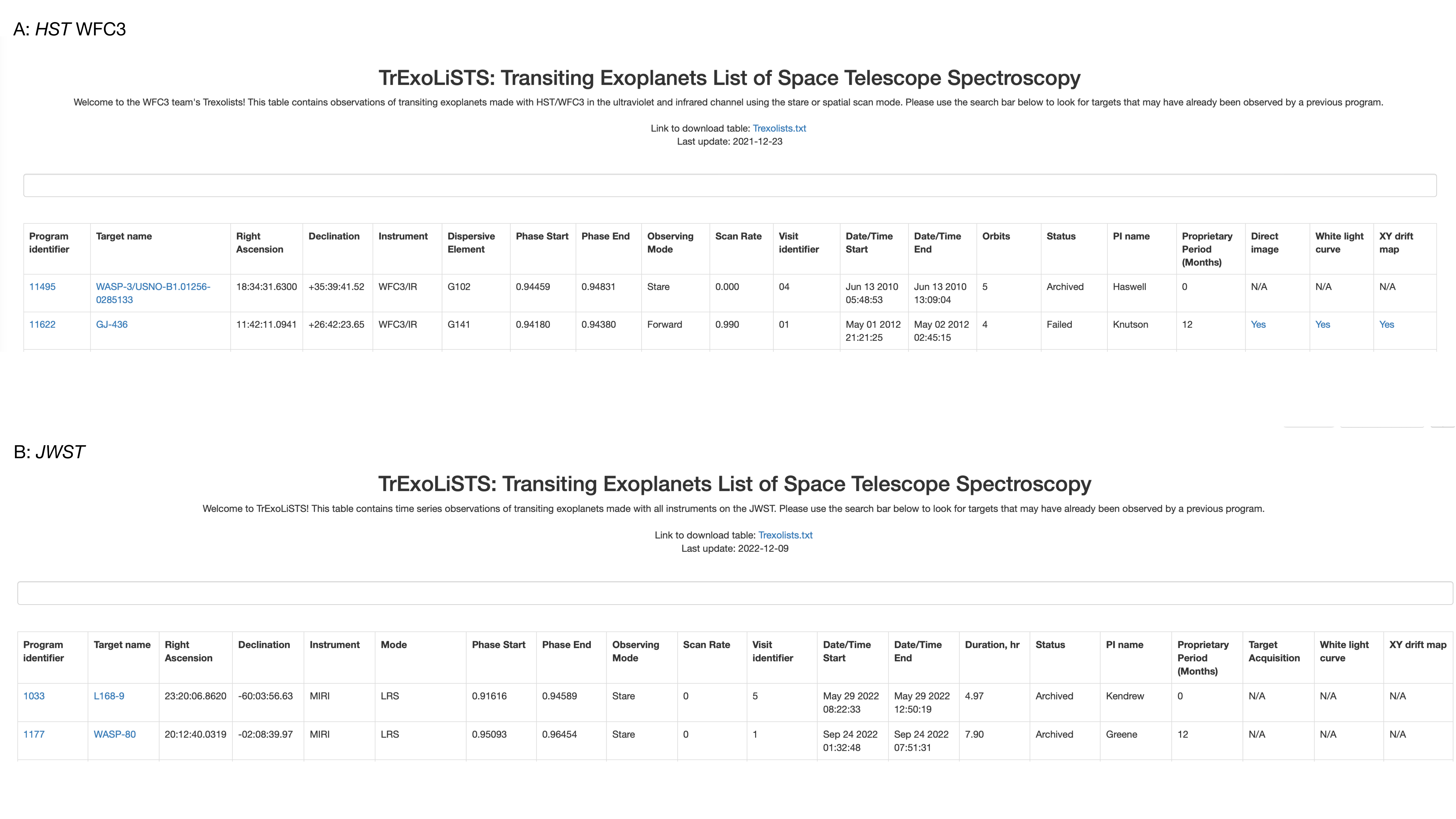}
\caption{Example of the current version of the TREXOLISTS catalogue for A:\,{\it{HST}} WFC3 and B:\,{\it{JWST}}. 
\label{fig:fig1}}
\end{figure}

TREXOLISTS is currently available at two separate addresses for each observatory:
 \begin{itemize}
\item{{\it{HST}} WFC3: \url{https://www.stsci.edu/~WFC3/trexolists/trexolists.html}}
\item{{\it{JWST}}: \url{https://www.stsci.edu/~nnikolov/TrExoLiSTS/JWST/trexolists.html}}
\end{itemize}
Figure\,\ref{fig:fig1} visualizes the look of the top portions of the {\it{HST}} WFC3 and {\it{JWST}} webpages.

\section{Conclusions and future outlook} \label{sec:floats}
TREXOLISTS has been designed with the goal of facilitating users during proposal preparation of exoplanet atmospheric programs that use {\it{HST}}, {\it{JWST}} and other missions. The catalog can be employed to identify target lists, which can then facilitate the craftsmanship of complementary targets and wavelength coverage for comprehensive atmospheric characterization of transiting exoplanets. Future improvements to the webpage will include: merging the {\it{HST}} and {\it{JWST}} catalogs under a single webpage, expanding program query to other {\it{HST}} instruments and reference for the quality of {\it{JWST}} visits.

\begin{acknowledgments}
We thank Jennifer Medina, Harish Khandrika, Kevin Stevenson, Peter McCullough, Sylvia Baggett, Brett Blacker, Karla Peterson, Bryan Hilbert and Brian Brooks for fruitful discussions during this project.
\end{acknowledgments}

%

\vspace{5mm}
\facilities{HST(WFC3), JWST(NIRCam, NIRSpec, NIRISS, MIRI)}


\software{IDL, Python}



\bibliography{nnbib}{}
\bibliographystyle{aasjournal}



\end{document}